\documentstyle[aps,multicol,epsfig]{revtex}
\begin{document}
\draft
\title{
Coulomb Blockade and Coherent Single-Cooper-Pair Tunneling\\ 
in Single Josephson Junctions }
\author{Michio Watanabe\footnote{Present address: Semiconductors 
Laboratory, RIKEN (The Institute of Physical and Chemical Research), 
2-1 Hirosawa, Wako-shi, Saitama 351-0198, Japan. 
(michio@postman.riken.go.jp)} 
and David B. Haviland} 
\address{Nanostructure Physics, 
The Royal Institute of Technology (KTH), 
Lindstedtsv\"agen 24, SE-100 44 Stockholm, Sweden}
\date{Received 9 January 2001}
\maketitle
\begin{abstract}
We have measured the current-voltage characteristics 
of small-capacitance single Josephson junctions 
at low temperatures ($T\leq0.04$~K),  
where the strength of the coupling 
between the single junction and the electromagnetic environment 
was controlled with one-dimensional arrays of dc SQUIDs.  
We have clearly observed Coulomb blockade of Cooper-pair tunneling 
and even a region of negative differential resistance, 
when the zero-bias resistance of the SQUID arrays is much higher 
than the quantum resistance $h/e^2\approx26$~k$\Omega$.  
The negative differential resistance is evidence 
of coherent single-Cooper-pair tunneling 
in the single Josephson junction.  

\end{abstract}

\begin{multicols}{2}

Small-capacitance superconducting tunnel junctions provide 
a novel system for studying 
the interplay between the Josephson phase 
and the charge on the junction electrode.  
These are quantum mechanically conjugate variables, 
and their behavior is influenced by dissipation, 
such as discrete tunneling of quasi-particles, 
coupling to an electromagnetic environment, etc.
The simplest and the most fundamental example is the single junction.  
Current-voltage ($I$-$V$) characteristics of single junctions have 
been the subject of extensive theoretical investigations~\cite{Ave91,Sch90}.  
Experimentally, however, the observation of charging effects 
such as ``Coulomb blockade" has been considered to be extremely difficult  
in single junctions because a high-impedance environment is necessary, 
and special care should be taken with the measurement leads~\cite{Ave91}.  
For this reason, thin-film resistors~\cite{Hav91} and tunnel-junction 
arrays~\cite{Gee90,Shi97} were employed for the leads, 
and an increase of differential resistance around $V=0$ was reported.  

We use one-dimensional arrays of 
dc superconducting quantum interference devices (SQUIDs) for the leads.  
The advantage of this SQUID configuration is that 
the effective impedance of array can be 
varied {\it in situ} by applying an external magnetic field 
perpendicular to the substrate.  
Thus, the zero-bias resistance of the SQUID arrays 
at low temperatures can be varied over several orders of magnitude.  
This phenomenon has been extensively studied in terms of 
the superconductor-insulator transition~\cite{Hav00}.  
The single junction in our samples, on the other hand, 
does not have a SQUID configuration, 
and therefore its parameters are practically independent 
of the external magnetic field.  
This enables us to study the {\it same} single junction in different 
environments.  This type of experiment has not yet been reported 
to the best of our knowledge.   
We show that the $I$-$V$ curve of the single junction is indeed sensitive 
to the state of the environment.  Furthermore, we can induce 
a transition to a Coulomb blockade of the single junction 
when the zero-bias resistance of the SQUID arrays is much higher 
than the quantum resistance $R_K\equiv h/e^2\approx26$~k$\Omega$.  

In addition to Coulomb blockade, we have clearly observed a region 
of negative differential resistance in the $I$-$V$ curve.  
Negative differential resistance has been reported 
in one-dimensional arrays~\cite{Hav00} and in two-dimensional 
arrays~\cite{Gee89}, however, clear observation in single junctions 
has not yet been reported in the literature.  

\begin{figure}
\centerline{
\psfig{
file=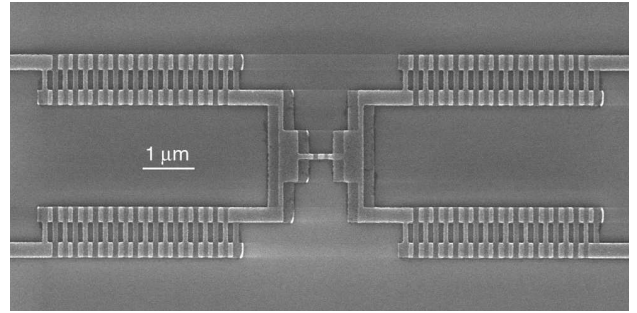,
width=0.95\columnwidth,
bbllx=7pt,
bblly=3pt,
bburx=583pt,
bbury=291pt,
clip=,
angle=0}
}
\caption{ 
A scanning electron micrograph of a sample.  
}
\label{fig:SEM}
\end{figure}

In the context of the theory of current-biased 
single Josephson junctions ~\cite{Ave91,Sch90},
the negative differential resistance appears 
as a result of coherent tunneling of single Cooper pairs.  
The onset of the coherent tunneling is characterized by 
the local voltage maximum in the low-current part 
of the $I$-$V$ curve, or blockade voltage $V_b$.  
The value of $V_b$ is suppressed as the ratio $E_J/E_C$ is increased.
Here, $E_J$ is the Josephson energy and $E_C\equiv e^2/2C$, where 
$C$ is the capacitance of the single junction.  
We show that the $E_J/E_C$ dependence of our measured $V_b$ 
is consistent with the theoretical prediction.  

A scanning electron micrograph of an Al/Al$_2$O$_3$/Al sample 
is shown in Fig.~\ref{fig:SEM}.  
A single junction with the area of 0.1$\times$0.1~$\mu$m$^2$ is in the 
center of Fig.~\ref{fig:SEM}.  On each side of the single junction there 
are two leads enabling four-point measurements of the single junction.  
A part of each lead close to the single junction consists 
of an array of dc SQUIDs.  
The area of each junction in the leads is 0.3$\times$0.1~$\mu$m$^2$ 
and the effective area of the SQUID loop is 0.7$\times$0.2~$\mu$m$^2$.  
All of the samples have the same configuration, except for 
the number $N$ of junction pairs in each lead.  
In Fig.~\ref{fig:SEM}, $N=17$ is shown,  
however, we measured the samples with larger $N$, up to 225.  
The tunnel junctions are fabricated on a SiO$_2$ substrate 
using electron-beam lithography and a double-angle-evaporation 
technique~\cite{Hav96}.  
The oxidation conditions, or the thickness of the Al$_2$O$_3$ layer, 
determines the tunnel resistance of the junctions.  
The samples are characterized by the normal-state resistance $R_n$ 
of the single junction,   the normal-state resistance $r_n$ 
per junction pair of the SQUID arrays, and $N$.  

\begin{figure}
\centerline{
\psfig{
file=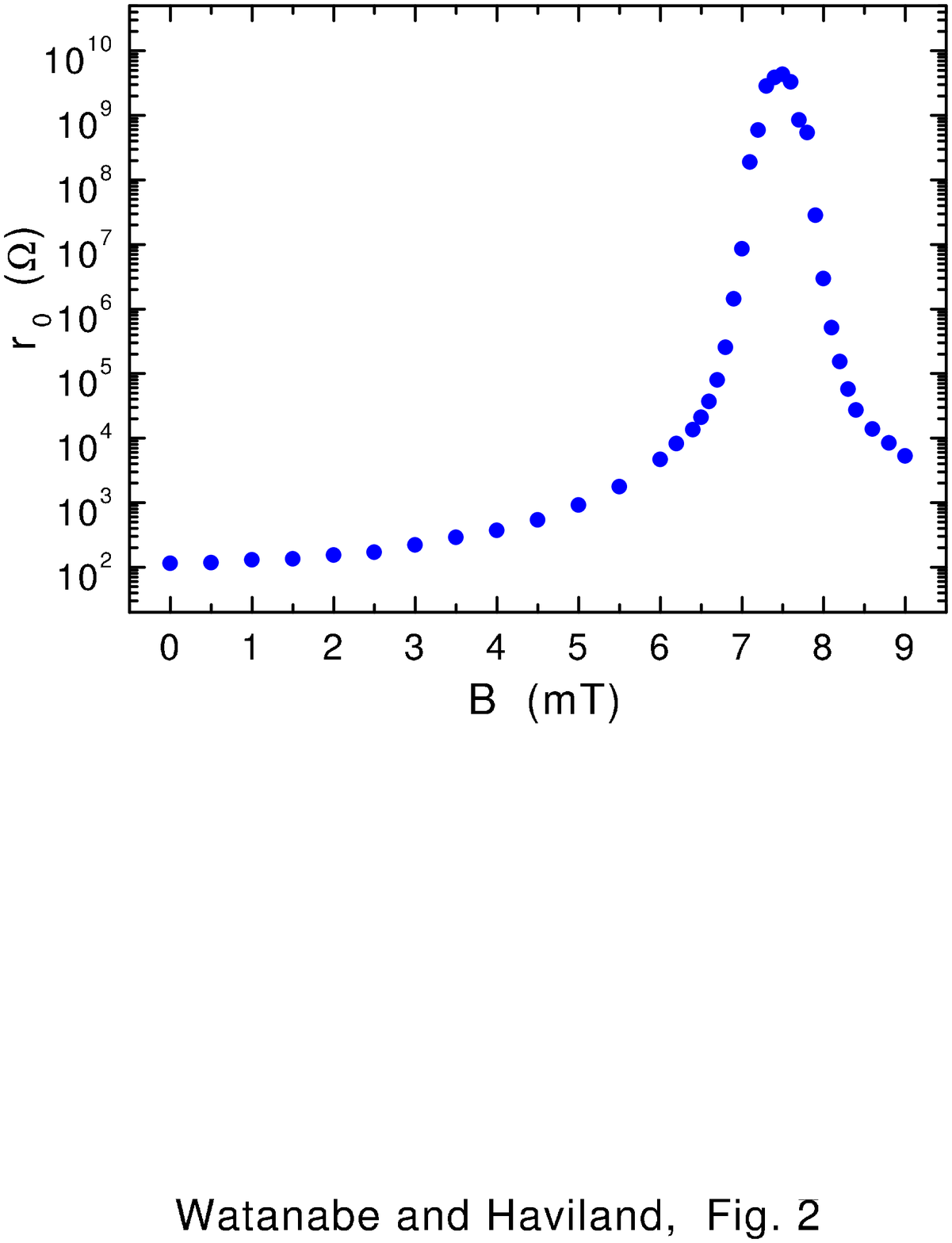,
width=0.85\columnwidth,
bbllx=22pt,
bblly=345pt,
bburx=555pt,
bbury=763pt,
clip=,
angle=0}
}
\caption{ 
Zero-bias resistance of two SQUID-array leads 
connected in series for a sample 
with $r_n=1.1$~k$\Omega$ and $N=225$.   
}
\label{fig:R0leads}
\end{figure}

The samples were measured in a $^{3}$He-$^{4}$He dilution refrigerator 
mainly at the base temperature ($0.02-0.04$~K).  
The temperature was determined by measuring the resistance 
of a ruthenium-oxide thermometer fixed at the mixing chamber.  
To measure the resistance of the thermometer, we used an ac resistance 
bridge (RV-Elekroniikka, AVS-47).  
The samples were placed inside a copper RF-tight box which was
thermally connected to the mixing chamber.  
All the leads entering the RF-tight box were low-pass filtered 
by 1~m of Thermocoax cable~\cite{Zor95}.  
We measured the $I$-$V$ curve of the single junction in a four-point 
configuration, where the potential difference was measured 
through one pair of SQUID-array leads with a high-input-impedance 
instrumentation amplifier based on two operational amplifiers 
(Burr-Brown OPA111BM) and a preamplifier [Stanford Research Systems 
(SRS) SR560].  Through the other pair of SQUID-array leads, the bias 
was applied, and the current was measured with a current 
preamplifier (SRS SR570).  
When the voltage drop at the SQUID arrays 
was much larger than that at the single junction, 
the single junction was practically current biased.  
The SQUID arrays could be measured in a two-point configuration 
(same current and voltage leads) on the same side of the single 
junction.  Note that the two arrays are connected in series 
and that current does not flow through the single junction.  
In order to obtain $R_n$ and $r_n$, we measured at $T=2-4$~K 
(above the superconducting transition temperature of Al).  

The effective Josephson energy between adjacent islands 
in the SQUID arrays is proportional to $|\cos(\pi BA/\Phi_0)|$,    
where $B$ is the external magnetic field applied perpendicular 
to the substrate, $A$ is the effective area of the SQUID loop, 
and $\Phi_0= h/2e= 2\times10^{-15}$~Wb 
is the superconducting flux quantum.  
Figure~\ref{fig:R0leads} shows the zero-bias resistance $r_0$
as a function of $B$, of two SQUID arrays on the same side 
of the single junction. 
The arrays have $r_n=1.1$~k$\Omega$ and $N=225$.   
In our samples with $A=0.14$~$\mu$m$^2$, $BA/\Phi_0$ 
becomes $1/2$ at $B=7.4$~mT.  Thus, $r_0$ should have the first 
local maximum at 7.4~mT, as observed in Fig.~\ref{fig:R0leads}.  
The value of $r_0$ at the local maximum is more than 10$^7$ times 
larger than that at $B=0$.  This means that we can tune 
the electromagnetic environment for the single junction 
over a wide range.  
Henceforth we indicate the magnetic field 
as the frustration $f\equiv BA/\Phi_0$.   

\begin{figure}
\centerline{
\psfig{file=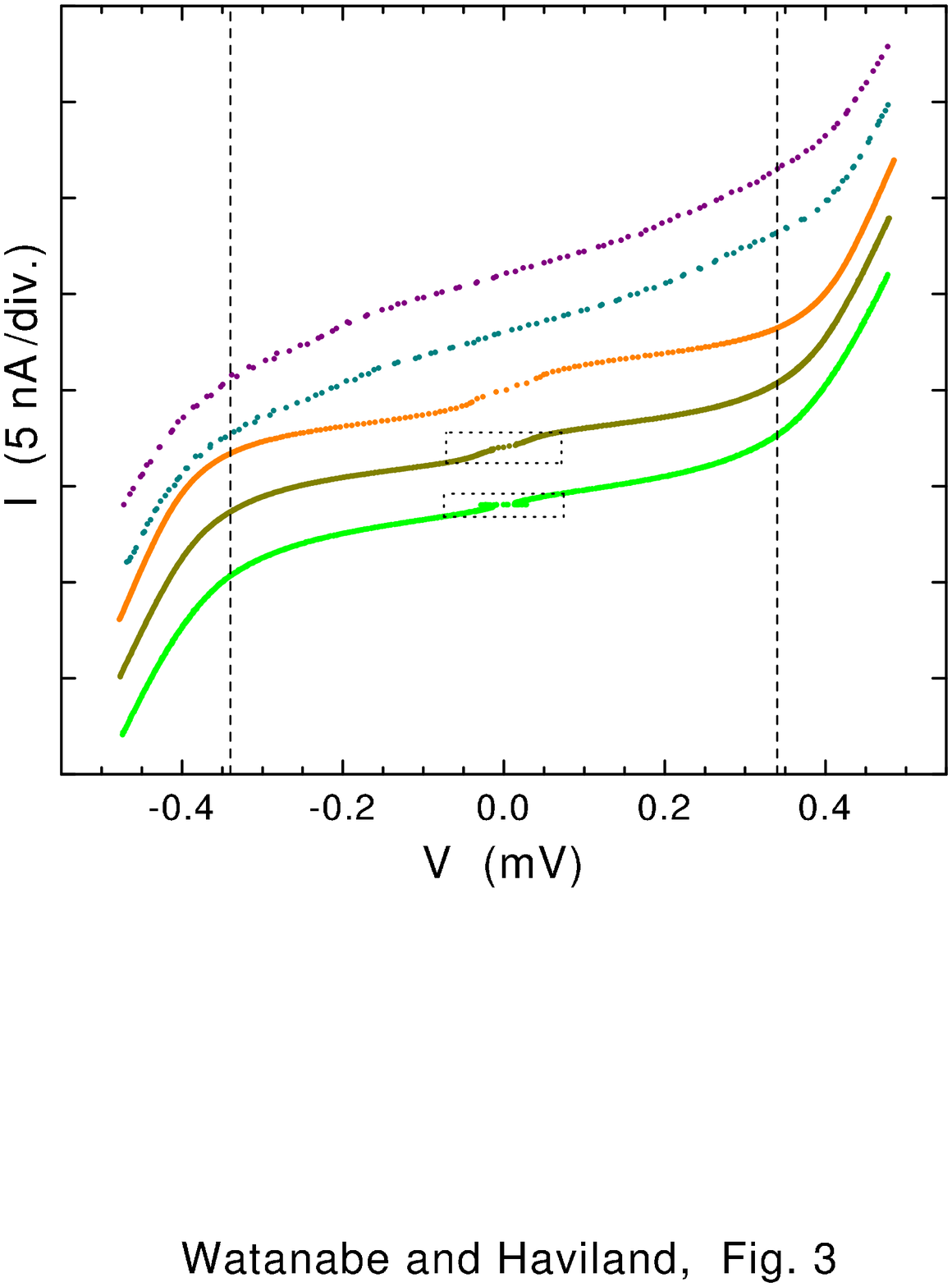,
width=0.85\columnwidth,
bbllx=37pt,
bblly=255pt,
bburx=561pt,
bbury=755pt,clip=,
angle=0}
}
\caption{ 
Current-voltage characteristics of the single junction  
for the sample with $R_n=17$~k$\Omega$, 
$r_n=1.4$~k$\Omega$, and $N=65$.  
From top to bottom, the frustration $f\equiv BA/\Phi_0$ 
is 0, 0.22, 0.41, 0.45, and 0.49, respectively.  
The origin of the current axis is offset 
for each curve for clarity.  
The vertical broken lines and the boxes with dotted lines 
represent the superconducting energy gap $\pm2\Delta$ 
and the region that will be closed up 
in Fig.~\ref{fig:21A4SJn}, respectively.    
}
\label{fig:21A4SJw}
\end{figure}

\begin{figure}
\centerline{
\psfig{file=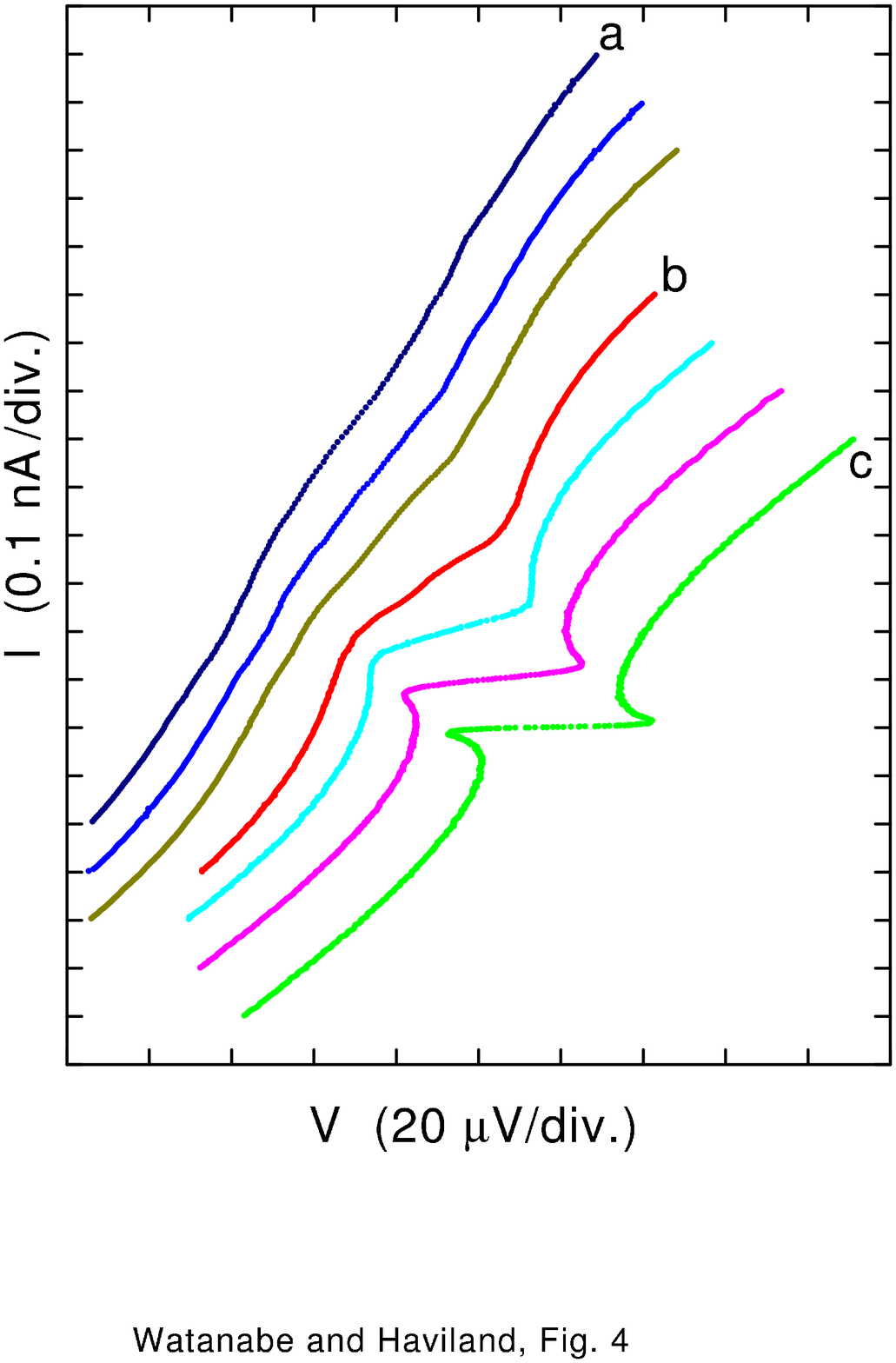,
width=0.85\columnwidth,
bbllx=37pt,
bblly=140pt,
bburx=541pt,
bbury=803pt,clip=, 
angle=0}
}
\caption{ 
Current-voltage ($I$-$V$) curves of the single junction  
for the same sample as in Fig.~\ref{fig:21A4SJw}.  
From top left to bottom right, the frustration $f$ 
is increased from 0.43 to 0.49 in steps of 0.01.  
The origin of each curve is offset for clarity.  
For the labeled curves, the $I$-$V$ characteristics 
of the leads at the same $f$ are shown in Fig.~\ref{fig:21A4leads}
}
\label{fig:21A4SJn}
\end{figure}

\begin{figure}
\centerline{
\psfig{file=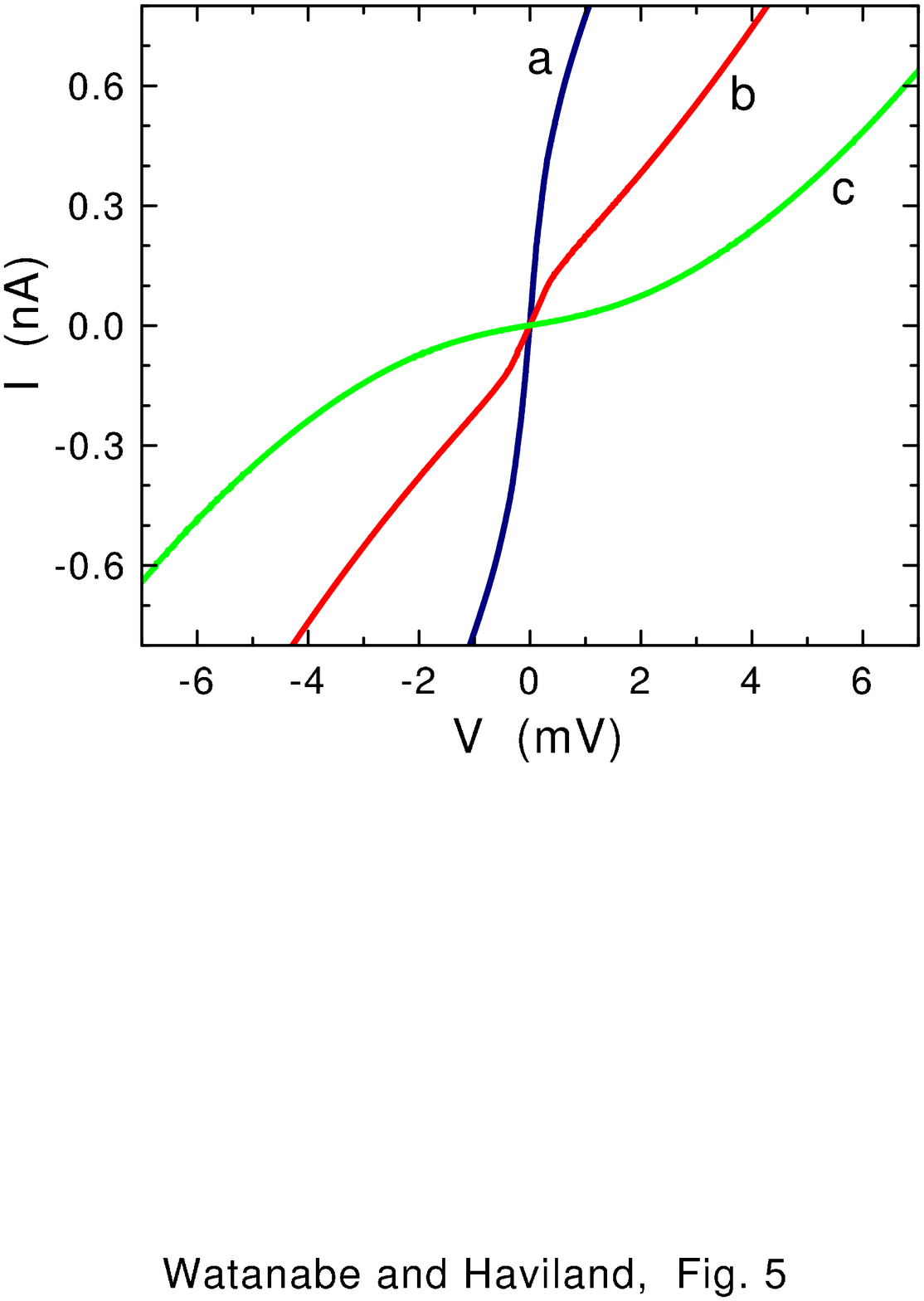,
width=0.85\columnwidth,
bbllx=24pt,
bblly=328pt,
bburx=557pt,
bbury=784pt,clip=,
angle=0}
}
\caption{ 
Current-voltage curves of the two SQUID-array leads 
connected in series for the same sample as in Figs~\ref{fig:21A4SJw} 
and \ref{fig:21A4SJn}.  From a to c, the frustration 
is 0.43, 0.46, and 0.49, respectively.   
}
\label{fig:21A4leads}
\end{figure}

In Figs.~\ref{fig:21A4SJw}--\ref{fig:21A4leads}, 
we show some results on a sample with 
$R_n=17$~k$\Omega$, $r_n=1.4$~k$\Omega$, and 
$N=65$.  The $I$-$V$ curves for the single junction 
at several frustrations are shown in Fig.~\ref{fig:21A4SJw}.  
The vertical broken lines represent the superconducting 
energy gap $\pm2\Delta$, which is $\pm0.34$~meV 
for Al~\cite{Kit96}.  
The features within the gap are dependent on $f$ 
especially in the vicinity of $V=0$.  
A close-up of the region denoted by the boxes with dotted lines 
is shown in Fig.~\ref{fig:21A4SJn}.  
As $f$ is varied, the single-junction $I$-$V$ curve shown 
in Fig.~\ref{fig:21A4SJn} develops a Coulomb blockade.  
We reemphasize that the Josephson energy of the single junction 
is independent of $f$, because it does not have a SQUID configuration 
and the field $f\Phi_0/A$ applied here is much 
smaller than the critical field for Al films ($\approx0.1$~T).  
The electromagnetic environment for the single junction (the SQUID array), 
however, is strongly varied with $f$.  Figures~\ref{fig:21A4SJw}  
and \ref{fig:21A4SJn} demonstrate that the low-bias region 
of the single-junction $I$-$V$ curve is indeed sensitive to the environment.  

The $I$-$V$ curves of the two SQUID-array leads connected in series 
at $f=0.43$ ($r_0=0.61$~M$\Omega$), 0.46 ($r_0=3.2$~M$\Omega$), 
and 0.49 ($r_0=43$~M$\Omega$) are shown in Fig.~\ref{fig:21A4leads}.  
The $I$-$V$ curves of the leads are nonlinear, and in general the SQUID 
array cannot be described by a liner impedance model~\cite{Hav00}.  
However, we may characterize the environment by their $r_0$.  
Coulomb blockade is visible only when $r_0\gg R_K$, 
which is consistent with the theoretical 
conditions for the clear observation of Coulomb blockade 
in single junctions~\cite{Ing92}.  
For an arbitrary linear environment characterized by $Z_e(\omega)$, 
$\mbox{Re}[Z_e(\omega)]\gg R_K$ is required for the Coulomb blockade 
of single-electron tunneling and $\mbox{Re}[Z_e(\omega)]\gg R_K/4$ 
for that of Cooper-pair tunneling~\cite{Ing92}.  
However, as seen in Fig.~\ref{fig:21A4leads}, 
our SQUID arrays are a nonlinear environment.  
It is interesting to note that at $f=0.46$ (labeled ``b"), the $I$-$V$ curve 
of the leads is still ``Josephson-like" (differential resistance is 
lower around $V=0$), while that of the single junction is already 
``Coulomb-blockade-like".  
This feature becomes more distinct in samples with larger $N$.  

The region of negative differential resistance seen 
in Fig.~\ref{fig:21A4SJn} when Coulomb blockade is well developed, 
is related to coherent tunneling of single Cooper pairs 
according to the theory~\cite{Ave91,Sch90} 
of a current-biased single Josephson junction in an environment 
with sufficiently high impedance.  
For sufficiently low current, the electrical conduction 
of single Josephson junctions is governed by stochastic 
quasi-particle tunneling, and the $I$-$V$ characteristic 
is highly resistive.  As the current is increased, 
coherent single-Cooper-pair tunneling, or ``Bloch oscillation" 
dominates, decreasing the mean voltage.  
As a result, the $I$-$V$ curve has a region of negative differential 
resistance, or ``back-bending" in the low-current part. 

Theoretically, the value of the local voltage maximum, or blockade 
voltage $V_b$, is a function of the ratio $E_J/E_C$, and given by   
\begin{equation}
V_b \approx 
\left\{
\begin{array}{ll} 
0.25\,e/C\;\; & \mbox{for $E_J/E_C \ll 1$,} \\
\delta_0/e & \mbox{for $E_J/E_C \gg 1$,}
\end{array}
\right.
\end{equation}
as $T\to 0$, where 
\begin{equation}
\delta_0 = \frac{\,e^2\,}{C}\,8\left(\frac{1}{2\pi^2}\right)^{\!1/4}\!
\left(\frac{E_J}{E_C}\right)^{\!3/4}\exp\left[-\left(8\,\frac{E_J}{E_C}
\right)^{\!1/2}\,\right]
\end{equation}
is the half width of the lowest energy band~\cite{Sch90}.   
When quasi-particle tunneling is neglected, 
$V_b$ becomes larger~\cite{Lik85},  
\begin{equation}
V_b \approx 
\left\{
\begin{array}{ll} 
e/C & \mbox{for $E_J/E_C \ll 1$,} \\
\pi\delta_0/e\;\; & \mbox{for $E_J/E_C \gg 1$.}
\end{array}
\right.
\end{equation}
The calculation of $V_b$ for arbitrary $E_J/E_C$ has also been 
done for the case of no quasi-particle tunneling~\cite{Zorin}.  

\begin{figure}
\centerline{
\psfig{file=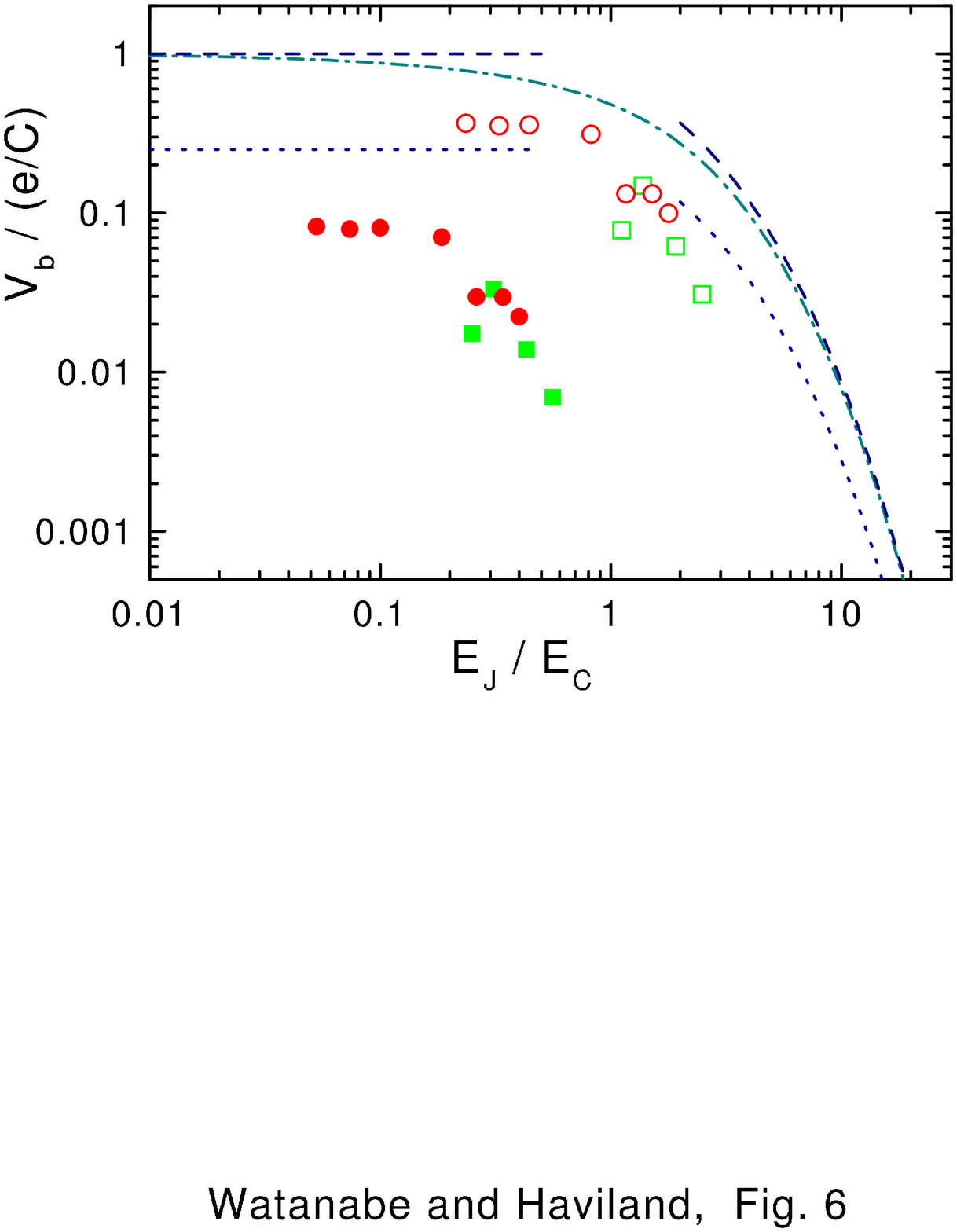,
width=0.85\columnwidth,
bbllx=30pt,
bblly=360pt,
bburx=568pt,
bbury=767pt,clip=,
angle=0}
}
\caption{
Blockade voltage $V_b$ divided by $e/C$ as a function 
of $E_J/E_C$.  From top to bottom, the curves represent 
the theoretical predictions of Refs.~[\ref{Lik85}], 
[\ref{Zorin}], and [\ref{Sch90}], respectively.  
For the calculation of the solid symbols (lower data set) 
45~fF/$\mu$m$^2$ is assumed, 
and for the open symbols (upper data set) 200~fF/$\mu$m$^2$.  
The solid boxes and the open boxes represent the samples 
with the nominal junction area of 0.01~$\mu$m$^2$ 
in Ref.~[\ref{Hav91}].
}
\label{fig:Vb}
\end{figure}

In Fig.~\ref{fig:Vb}, we compare the measured $V_b$ with the above 
predictions.  
We used $E_J=h\Delta/8e^2R_n$, and for $E_C$ we estimated $C$ 
from the junction area.  
A value of specific capacitance $c_s=45\pm5$~fF/$\mu$m$^2$~\cite{Lic89}, 
which was obtained for the junctions with $3\times28$~$\mu$m$^2$ 
and $7\times54$~$\mu$m$^2$, has been frequently 
employed~\cite{Hav91,Shi97,Hav00}.  
Uncertainty in $c_s$, however, seems to be much larger 
when the junction area is on the order of 0.01~$\mu$m$^2$ or smaller.  

In the following experimental literature on small tunnel junctions, 
$C$ was estimated from the offset voltage in the normal-state $I$-$V$ curve.    
Fulton and Dolan measured samples with three junctions 
that share a common electrode, and obtained 
$0.20-0.23$~fF for ($0.03\pm0.01$~$\mu$m)$^2\times3$~\cite{Ful87}, 
i.e., $c_s=42-192$~fF/$\mu$m$^2$.  
Geerligs {\it et al.} reported $c_s\approx 110$~fF/$\mu$m$^2$ 
for two-dimensional ($190\times60$) junction arrays with 
the areas of 0.01 or 0.04~$\mu$m$^2$~\cite{Gee89}.  
More recently, Penttil\"a {\it et al.} studied resistively shunted 
single Josephson junctions with the area of 
$0.15\times0.15$~$\mu$m$^2$~\cite{Pen99}.  
The estimated $C$ of their eight samples ranged between 0.8 and 6.6~fF, 
or $c_s=36-293$~fF/$\mu$m$^2$.  

Motivated by the above survey, in Fig.~\ref{fig:Vb} we plot 
the experimental $V_b$ vs $E_J/E_C$, where two different values of 
$c_s$ have been used to determine $C$.  The lower data set uses
45~fF/$\mu$m$^2$ and the upper data set 200~fF/$\mu$m$^2$.  
The data from Ref.~\cite{Hav91} for the samples with the same nominal 
junction area ($0.1\times0.1$~$\mu$m$^2$) as in this work are also plotted.  
The measured $V_b$ agrees with the theoretical predictions qualitatively.  
We also find a quantitative agreement when $c_s\approx 200$~fF/$\mu$m$^2$ 
is taken for 0.01~$\mu$m$^2$.  
This apparently large junction capacitance may be partly explained 
by distributed capacitance of the SQUID arrays.  

In summary, we have studied the current-voltage characteristics 
of single Josephson junctions biased with SQUID arrays.  
We have demonstrated that the Coulomb blockade in the single junction 
is indeed sensitive to the state of the environment.  
The region of negative differential resistance has been 
clearly observed in the $I$-$V$ curve of the single junction, 
when the zero-bias resistance of the SQUID arrays is much higher 
than $R_K$, which is clear evidence of coherent single-Cooper-pair 
tunneling in the single Josephson junction.  

We are grateful to T. Kato and R. L. Kautz for fruitful discussions.
This work was supported by Swedish NFR.  
M. W. would like to thank Japan Society for the Promotion of 
Science (JSPS) and the Swedish Institute (SI) for financial supports.  

\end{multicols}

\begin{references}
\bibitem{Ave91}D. V. Averin and K. K. Likharev, in {\it Mesoscopic Phenomena 
	in Solids}, edited by B. L. Altshuler, P. A. Lee and R. A. Webb 
	(Elsevier Science Publishers B. V., Amsterdam, 1991), chap.~6.  
\bibitem{Sch90}G. Sch\"on and A. D. Zaikin, Phys. Reports {\bf 198}, 
	237 (1990).  \label{Sch90}
\bibitem{Hav91}D. B. Haviland, L. S. Kuzmin, P. Delsing, K. K. Likharev, 
	and T. Claeson, Z. Phys. B {\bf 85}, 339 (1991).  \label{Hav91}
\bibitem{Gee90}L. J. Geerligs, Ph.D Thesis, Delft Univ. of Tech. (1990). 
\bibitem{Shi97}Y. Shimazu, T. Yamagata, S. Ikehata, and S. Kobayashi, 
	J. Phys. Soc. Jpn. {\bf 66}, 1409 (1997).  
\bibitem{Hav00}D. B. Haviland, K. Andersson, and P. \r{A}gren, 
	J. Low Temp. Phys. {\bf 118}, 733 (2000).  
\bibitem{Hav96}D. B. Haviland, S. H. M. Persson, P. Delsing, and 
	C. D. Chen, J. Vac. Sci. Technol. A {\bf 14}, 1839 (1996).  
\bibitem{Zor95}A. B. Zorin, Rev. Sci. Instrum. {\bf 66}, 4296 (1995).  
\bibitem{Gee89}L. J. Geerligs, M. Peters, L. E. M. de Groot, A. Verbruggen, 
	and J. E. Mooij, Phys. Rev. Lett. {\bf 63}, 326 (1989). 
\bibitem{Kit96}C. Kittel, {\it Introduction to Solid State Physics}, 7th ed. 
(John Wiley \& Sons, New York, 1996), p.~344.  
\bibitem{Ing92}G.-L. Ingold and Y. V. Nazarov, in {\it Single Charge 
Tunneling}, edited by H. Grabert and M. H. Devoret (Plenum Press, 
New York, 1992), chap.~2.   
\bibitem{Lik85}K. K. Likharev and A. B. Zorin, J. Low Temp. Phys. {\bf 59}, 
	347 (1985).  \label{Lik85}
\bibitem{Zorin}A. B. Zorin (unpublished).  \label{Zorin}
\bibitem{Lic89}A. W. Lichtenberger, C. P. McClay, R. J. Mattauch, 
	M. J. Feldman, S.-K. Pan, and A. R. Kerr, IEEE Trans. Magn. 
	{\bf 25}, 1247 (1989).  
\bibitem{Ful87}T. A. Fulton and G. J. Dolan, Phys. Rev. Lett. {\bf 59}, 
	109 (1987).  
\bibitem{Pen99}J. S. Penttil\"a, \"U. Parts, P. J. Hakonen, M. A. Paalanen, 
	and E. B. Sonin, Phys. Rev. Lett. {\bf 82}, 1004 (1999).  
\end{references}
\end{document}